# An OpenCL-based Monte Carlo dose calculation engine (oclMC) for coupled photon-electron transport

**Zhen Tian, Feng Shi, Michael Folkerts, Nan Qin, Steve B. Jiang, Xun Jia**

Department of Radiation Oncology, University of Texas Southwestern Medical Center, Dallas, TX 75390, USA

Emails:      zhen.tian@utsouthwestern.edu,      steve.jiang@utsouthwestern.edu, xun.jia@utsouthwestern.edu

Monte Carlo (MC) method has been recognized the most accurate dose calculation method for radiotherapy. However, its extremely long computation time impedes clinical applications. Recently, a lot of efforts have been made to realize fast MC dose calculation on GPUs. Nonetheless, most of the GPU-based MC dose engines were developed in NVidia's CUDA environment. This limits the code portability to other platforms, hindering the introduction of GPU-based MC simulations to clinical practice. The objective of this paper is to develop a fast cross-platform MC dose engine oclMC using OpenCL environment for external beam photon and electron radiotherapy in MeV energy range. Coupled photon-electron MC simulation was implemented with analogue simulations for photon transports and a Class II condensed history scheme for electron transports. To test the accuracy and efficiency of our dose engine oclMC, we compared dose calculation results of oclMC and gDPM, our previously developed GPU-based MC code, for a 15 MeV electron beam and a 6 MV photon beam on a homogenous water phantom, one slab phantom and one half-slab phantom. Satisfactory agreement was observed in all the cases. The average dose differences within 10% isodose line of the maximum dose were 0.48-0.53% for the electron beam cases and 0.15-0.17% for the photon beam cases. In terms of efficiency, our dose engine oclMC was 6-17% slower than gDPM when running both codes on the same NVidia TITAN card due to both different physics particle transport models and different computational environments between CUDA and OpenCL. The cross-platform portability was also validated by successfully running our new dose engine on a set of different compute devices including an Nvidia GPU card, two AMD GPU cards and an Intel CPU card using one or four cores. Computational efficiency among these platforms was compared.

**1. Introduction**

Dose calculation plays a central role in radiation therapy. Among many existing algorithms, Monte Carlo (MC) simulation method is commonly regarded as the ground truth due to its capability of accurately simulating a particle transport process and modeling simulation geometry (Rogers, 2006). Computational efficiency is a major issue preventing clinical adoptions of this novel method. Despite dramatically increased CPU speed and number of cores over the years, MC simulation codes are often slow for most routine applications and improving their efficiency is an active research topic. Recently, along with the boom in GPU-based high-performance computing, there has been a burst of researches regarding developing MC simulation tools on this platform (Pratx and Xing, 2011; Jia *et al.*, 2014). Substantial acceleration factors over conventional CPU-based MC tools have been reported by a number of research groups for a variety of particle types (photon, electron, and proton) in different energy ranges (Jia *et al.*, 2010; Jia *et al.*, 2011b; Hissoiny *et al.*, 2011; Jahnke *et al.*, 2012; Townson *et al.*, 2013; Bol *et al.*, 2012; Hissoiny *et al.*, 2012; Jia *et al.*, 2012; Yepes *et al.*, 2010; Badal and Badano, 2009).

To date, majority of GPU-enabled MC dose engines have been developed on NVidia GPU cards (Santa Clara, CA) under its Compute Unified Device Architecture (CUDA) (NVIDIA, 2011) due to the leading role of NVidia in GPU-based scientific computing. This fact holds not only for MC simulation problems, but for a wide spectrum of other applications in medical physics e.g. image processing (Jia *et al.*, 2011a; Jia *et al.*, 2011c; Gu *et al.*, 2010; Samant *et al.*, 2008), non-MC dose calculations, (Fujimoto *et al.*, 2011; Hissoiny *et al.*, 2009; Jacques *et al.*, 2010), and inverse treatment planning (Men *et al.*, 2010; Fei *et al.*, 2012). While great successes have been achieved in these projects, portability of those packages is a major issue, as they cannot be executed on GPU cards from other manufactures. To some extent, this fact has hindered the introduction of GPU technology into clinical practice, as radiotherapy vendors are probably reluctant to tie their products on a single hardware platform manufactured by a specific vendor. In addition, most of the existing packages were written in a GPU-specific language, e.g. CUDA. It is impossible to use then on the conventional CPU platform, leading to the concern of wasting the code development efforts.

Recently, Open Computing Language (OpenCL) was introduced into high-performance computing field. It enables a framework to write a program that executes across different platforms, including conventional CPUs, GPUs from different manufactures, and other processors. Developing versatile applications executable on different platforms potentially provides a solution to the portability problem and to eliminate the concern of wasting development efforts on the GPU platform. This will probably facilitate the wide adoption of GPU into clinical practice. Motivated by this fact, we have recently started developing a new MC dose engine for photon radiotherapy dose calculations in the OpenCL framework and this paper reports our progress towards this direction. To date, only a few applications in radiotherapy have been developed through OpenCL (Ammazzalorso *et al.*, 2014; Zhou *et al.*, 2012). We expect that our developments will help exploring the potential of this cross-platform computation framework for medical physics in radiotherapy.

It is also our objective to test the feasibility of achieving fast MC dose calculation using OpenCL. For this framework, being compatible with different hardware platforms also raised the concern of reduced computational efficiency, as it limits to what degree one can assume about the hardware structure and hence the room of tailoring a code for the underlying hardware. In the past a few years, several studies investigated this issue in a few example problems that are not radiotherapy related by comparing the efficiency of CUDA and OpenCL implementations (Su *et al.*, 2012; Kakimoto *et al.*, 2012; Fang *et al.*, 2011; Weber *et al.*, 2011; Habich *et al.*, 2013). While in most of the cases, CUDA won in speed, the advantages were sometimes not significant. It was also expected that the performance comparison depends on specific problem type and size. Hence, it is also an important problem to find out the speed that we can actually achieve in the MC dose calculation problem, which will help assessing the value of OpenCL for radiotherapy.

## 2. Methods and Materials

### 2.1 Physics

Our MC code, named oclMC, targets at performing dose calculations for external beam photon or electron beam therapy. It simulates coupled photon-electron transport in Mega eV energy range. The physics for MC simulation in this process is well known and here we only briefly present them.

We modeled a patient using a voxelized geometry. Each voxel was assigned to a density value and a material type, which can be obtained through a mapping method from a patient CT image (Schneider *et al.*, 2000). At present, a total number of 16 materials that are commonly used in radiotherapy were supported in oclMC such as water, ICRU lung, cortical bone, compact bone, adipose tissue etc. The physics data for each material were stored in a database, such as photon attenuations for different interaction channels, electron restricted stopping power, and macroscopic cross sections. A computer program was also developed to create physics data needed by oclMC based on a user-defined material composition. The photon data for each element were taken from NIST XCOM database, and that for electron were computed using analytical formula (Kawrakow *et al.*, 2011).

For photons in the MeV energy range, Compton scattering, pair production and photoelectric absorption were modeled in our simulation. Woodcock tracking method (Woodcock *et al.*, 1965) was employed to handle photon transport, which significantly increased the efficiency by eliminating calculations of voxel boundary crossing in a heterogeneous phantom. At a Compton scattering event, the scattered angle was sampled from a Klein-Nishina differential cross section, and the properties of the secondary electron followed from the kinematics. For a photoelectric event, the photon was absorbed locally. For a pair production event, a rough approximation was made, since this event is important only at the high end of the energy range and only for materials with high atomic numbers. The generated electrons and positrons were along the same direction as the incoming photon and the photon energy, after subtracting the mass energies of the particles, was randomly distributed to them.

Electron transport was simulated in a standard Class II condensed history scheme. An electron moved forward in a step-by-step fashion. In each step, it was only allowed to propagate for a distance $d = \min[d_{vox}, d_{maxs}, d_{maxe}, d_{event}]$, where $d_{vox}$ is the distance to the next voxel boundary along the particle direction. $d_{maxs}$ and $d_{maxe}$ are user defined maximum step size to constrain that the electron cannot move too far with a large energy decrease within a step. $d_{event}$ is the distance to the next hard event (Moller or Bremsstrahlung interactions). $d_{event}$ was sampled based on energy-dependent total cross section data for these two interactions. Woodcock method was used again to account for the variation of cross section within the step due to electron energy loss (Woodcock *et al.*, 1965; Kawrakow, 2000). At a Moller event, the kinetic energies of the two secondary electrons were sampled using cross section differential in the energy domain. Then their directions were determined based on kinematics. For a Bremsstrahlung event, a secondary photon was generated by sampling its energy according to the energy differential cross section data. Its deflection angle was approximated to be the angle with the maximum differential cross section. The direction of the incoming electron was unchanged. Within a step, a soft interaction was simulated using the random hinge method, where the multiple Coulomb scattering angle was determined using the method developed by Kawrakow (Kawrakow, 2000). Once an electron moved forward for a step, its energy loss was calculated using the restricted stopping power via the continuously slowing down approximation (CSDA), which was deposit to the corresponding voxel. A positron was transported in the same way as an electron, except that two annihilation photons traveling to two opposite directions were generated at the end of its track.

*2.2 OpenCL Implementation*

*2.2.1 Parallelization scheme*

In OpenCL, parallel computation is achieved by a number of work-items. These work-items are grouped into a number of work-groups and are executed by the available parallel processing units, e.g. multiprocessors on an NVidia GPU.

The main workflow of our simulation is illustrated in the algorithm shown in Table 1. The first step initialized all the physics and patient data. These data were loaded to GPU to be accessible to the work-items during simulation. Detailed usage of GPU memory in the OpenCL framework will be presented later. After that, there were two loops. First, we formed a group of energy bins. The dose calculation looped over these bins and sequentially performed simulations for source particles in each bin. This strategy ensured that all source particles being simulated concurrently had energies in a certain range. It helps to reduce loss of efficiency due to the existence of a GPU thread with a much longer computation time than the rest, which typically corresponds to a particle with a higher energy.

**Table 1**. Algorithm to illustrate the workflow of our MC simulation.

| 1. | Initialize |
|----|-----------|
| 2. | Loop over all energy bins |
| 3. | Loop till all the particles in the selected energy bin are simulated |
| 4. | Set a number of particles of same type (up to *M*) to be simulated |
| 5. | Transport these particles and put secondary particles in stacks |
| 6. | End |
| 7. | End |

The second loop was designed to control the types of particles being simulated concurrently. In a GPU-based dose calculation, a user specified the number of work-items *M* that are simultaneously launched. These work-items transport *M* particles at the same time, one for a particle. It is of importance to design a scheme to ensure that all the particles concurrently simulated are of the same type to alleviate GPU thread divergence problem (Hissoiny *et al.*, 2011; Jia *et al.*, 2011b). As such, two stacks were allocated in our code to separately hold secondary photons and electrons that are produced during particle transport. Within each iteration step of this loop, the dose engine first set particles to be simulated. These particles may either be from one of the two stacks, or be generated from the source. The decision was made such that we always chose the group that had the largest number of particles to be simulated among these three choices to better utilize the GPU's parallelization power. After that, these particles were transported simultaneously and the secondary particles generated during the transport simulation were put into corresponding stacks according to their types. It is worth mentioning that a too large value of *M* reduces efficiency, as there are not enough hardware resources, while a small *M* value underutilizes the GPU's computing power. In our implementation, we manually adjusted the value of *M* for each hardware platform for the best performance.

### 2.2.2 Memory management

Memory is another important aspect on GPU computation, which deserves detailed descriptions. There are four types of GPU memory in OpenCL: (1) Global memory, which is accessible to all GPU work-items. OpenCL API defines a memory object pointing to the global memory. A memory object can be either a buffer object to store a group of one-dimensional elements or an image object to store two- or three-dimensional elements. (2) Constant memory, which is a region of global memory that remains constant during the execution of a kernel function. (3) Local memory, which is a memory region associated with a work-group and accessible only by those work-items in that work-group. (4) Private memory, which is private to a work-item and not visible to any other work-items.

There were different types of data involved in our dose engine. They were allocated on different types of GPU memories depending on their specific usages during dose calculations. First, we chose to store the physics data, e.g. cross section and stopping power as a function of energy, and phantom data, e.g. density and material type, in global memory as 2D/3D images through image objects. It is actually desirable to store these read-only data in GPU's constant memory for a fast access. However, the size of GPU constant memory is usually too small to hold these objects. The 2D/3D image objects are similar to texture memory in CUDA, allowing fast hardware interpolations using built-in sampler functions. Yet, the usage of an image object in OpenCL is not as flexible as the texture object in CUDA. For instance, each image object must have four channels in OpenCL, e.g. each pixel/voxel has four elements. Hence, we grouped every four sets of cross section data, e.g. those for the three photon interaction types and the total cross section data, and put them into a 2D image object with each channel corresponding to one set of the cross section data. The two dimensions in this 2D image are energy and material types. Similarly, multiple differential cross sections were grouped and stored in different channels of a 3D image object. This strategy helps better utilize GPU memory.

For those variables that were kept constant during dose calculations, e.g. voxel sizes and phantom resolutions, we defined them through macros. This is possible because OpenCL actually compiles kernel functions at the run-time. Since each kernel source code is a simple text, we developed a CPU program at the initialization stage of the package to modify the kernel source code, specifically, the macro definition part, to specify the value of these variables. At run time, the correct variable values were incorporated into kernel executions after code compilation.

The space to store the particles (up to *M*) that were transported simultaneously was allocated on GPU global memory. At the beginning of kernel execution to transport them, the particle information was read from global memory by each work-item into its own private memory. This eliminates repeated visits to the slow global memory during particle transport simulations. The two stacks to store the secondary photons or electrons generated during simulations were allocated on the GPU global memory as well. When a work-item attempted to put a particle into the stack, atomic operations were utilized to ensure no other work-items can interfere this process.

Finally, the dose counter to tally the dose deposited into the phantom was allocated on GPU global memory due to its large data size. Each work-item directly deposited dose into voxels during the particle transport simulations. Atomic addition operation was used again to ensure result integrity under possible simultaneous dose depositions to the same voxel from multiple work-items.

### 2.2.3 Random number generator

Unlike CUDA that has a CURAND (NVIDIA, 2010) library to provide random numbers, there is no such a random number generator in OpenCL. Hence, we implemented a pseudo-random number generator RANECU (James, 1990) on GPU. Each work-item hold its own random number seed. The initial random seed value for each work-item was generated by a random number generator on CPU with system clock as its initial seed. Since each OpenCL work-item uses its own random seed, the random number sequences generated by different work-items were assumed to be independent.

### 2.2 Materials

To investigate the performance of our OpenCL-based MC dose engine, three phantoms were used in our experiments. The first one was a homogenous water phantom and the second was a layered slab phantom consisting of four layers along $z$ direction, namely, water, bone and lung layers with a thickness of 2 cm for each of them and then a water layer for the rest thickness. The third phantom used in our experiments was a half-slab phantom. Its configuration was similar to the slab phantom but the bone and the lung layers covered half of the $xoy$ plane, shown in Figure 3(a). All of these three phantoms were 32×32×32 $cm^3$ cubes with a voxel size of 0.25×0.25×0.25 cm$^3$.

It was not the focus of this paper to accurately model a linear accelerator and generate source particles for patient dose calculations. Hence, we employed a simple point source to test the efficiency and accuracy of our dose engine. However, we have also left the interface for source generation open in this package. A user can either load source particles from a phase-space file or write a CPU/GPU function to generate particles using their own analytical source model.

For photon beam cases, the source had a 6MV energy spectrum. For electron beam cases, the source was 15 MeV mono-energetic. In both cases, the source normally impinged at the center of the phantoms on the $xoy$ plane with 100 cm source-to-surface distance (SSD). Field sizes were set to 2×2 cm$^2$ at the phantom surface. In our simulations, we set the absorption energies to 200 keV for electron and 50 keV for photon, and the cutoff energies of hard events for electron transport to be 200 keV and 50 keV for bremsstrahlung and Moller interactions, respectively. For comparison purpose, our previously developed gDPM was also launched. The same cut-off energies were used to ensure a fair comparison of computational efficiency. Dose calculation results in both packages were reported in a unit of MeV/g per source particle to make accuracy comparisons. The cross-platform portability of our OpenCL-based MC dose calculation package was demonstrated by running it on several platforms including an Nvidia GeForce GTX TITAN GPU card, an AMD radeon R9 290x GPU card, an AMD radeon HD 7500s GPU card and an Intel i7-3770 CPU using one thread or all the eight threads. The simulation time on different platforms were presented. Moreover, the time of gDPM and our new package on the Nvidia GPU card was also recorded for efficiency comparison.

## 3. Results

### 3.1 Accuracy Evaluations

The dose of the 2×2 $cm^2$ 15MeV electron beam in the homogeneous water phantom and the water-bone-lung-water slab phantom were calculated using both our OpenCL-based MC code oclMC and gDPM code for comparison, shown in Figure 1. The two rows represented the water phantom and the slab phantom, respectively. The two columns were depth-dose curves along the beam central axis and lateral dose profiles at depth 1 cm, 3 cm, 5 cm, 7 cm, respectively. We simulated $5 \times 10^6$ source electrons and the average relative uncertainty $\overline{\sigma_D/D}$ within 10% isodose line of the maximum dose was 0.52% for both the water phantom case and the slab phantom case. Here $\sigma_D$ denotes the uncertainty of the local dose $D$ at a voxel. The error bar sizes in Figure 1 were $2\sigma_D$ for the dose calculated by our dose engine oclMC. For the purpose of clarity, the error bars of the dose calculated by gDPM were not shown, which were of similar sizes to those of the OpenCL results. Good match between our oclMC dose and gDPM dose was observed in both phantom cases. The doses fell quickly to zero due to electron scattering and continuous energy loss.

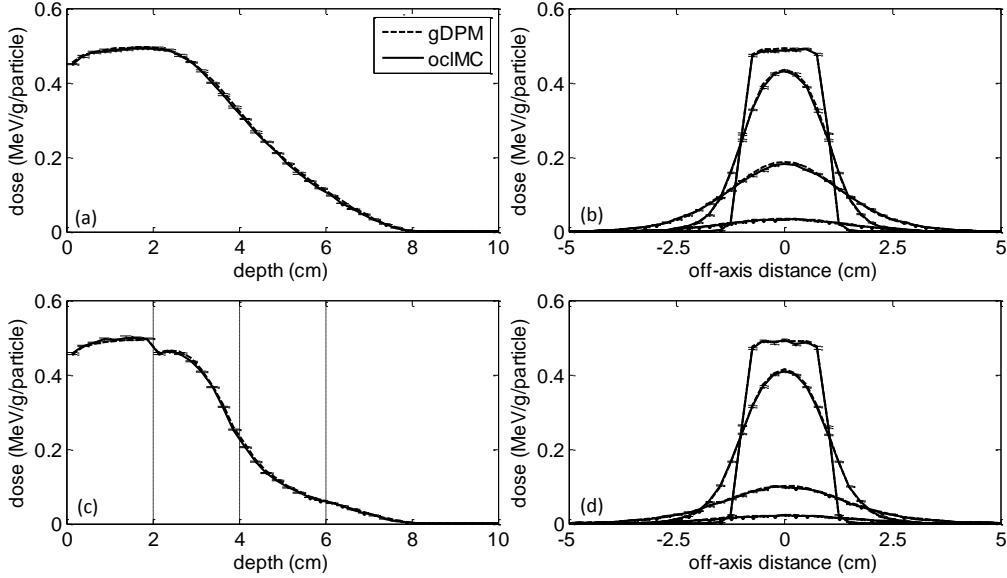

**Figure 1.** Depth-dose curves (left) and lateral dose profiles at different depths 1 cm, 3 cm, 5cm, 7cm (right) for a homogeneous water phantom (top) and a water-bone-lung-water phantom (bottom) with a $2 \times 2\ cm^2$, 15 MeV electron point source. Error bars show $2\sigma_D$ of dose calculated by our OpenCL-based dose engine.

We have also quantitatively evaluated the accuracy of our dose engine oclMC by calculating the dose difference compared to gDPM for each point within 10% isodose line. The results are presented in Table 1. The average difference was 0.48% relative to the maximum dose for the water phantom and 0.53% for the slab phantom. More than 98% of the calculation points have a dose difference within 1.67% and 1.61% for the water phantom and the slab phantom, respectively.

**Table 1.** Quantitative comparison results for electron and photon sources on different phantoms.

| Beam | No. of particles | phantom | $\overline{\sigma_D/D}$ (%) | | Average Difference (%) | 98%-Max Difference (%) |
|---|---|---|---|---|---|---|
| | | | gDPM | oclMC | | |
| 15MeV electron | $5 \times 10^6$ | water | 0.53 | 0.52 | 0.48 | 1.67 |
| | | slab | 0.54 | 0.52 | 0.53 | 1.61 |
| 6MV photon | $5 \times 10^8$ | water | 0.33 | 0.33 | 0.15 | 0.53 |
| | | slab | 0.33 | 0.33 | 0.16 | 0.67 |
| | | half-slab | 0.31 | 0.31 | 0.17 | 0.73 |

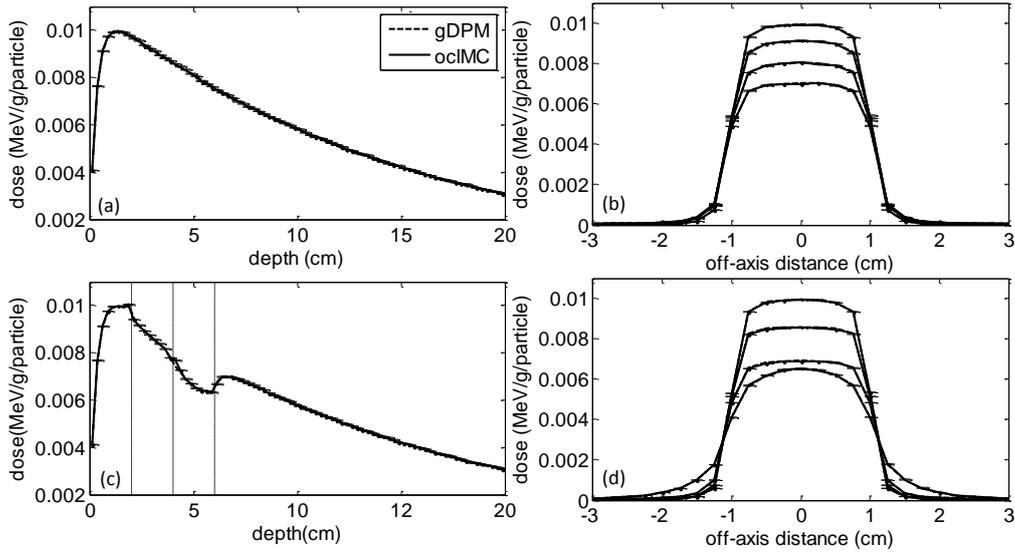

**Figure 2.** Depth-dose curves (left) and lateral dose profiles at different depths 1 cm, 3 cm, 5cm, 7cm (right) for a homogeneous water phantom (top) and a water-bone-lung-water phantom (bottom) with a 2×2 $cm^2$, 6 MV photon point source. Error bars show $2\sigma_D$ of dose calculated by our OpenCL-based dose engine.

The dose of the 2×2 $cm^2$ 6MV photon beam in the water phantom and the slab phantom were shown in the two rows of Figure 2, respectively. Similar to the electron results, the doses calculated by our OpenCL-based MC code were in a good agreement with those calculated by gDPM. The dose results for the slab phantom with a photon source were displayed in Figure 3, where the subfigure (a) showed the configuration of this half slab phantom with four dash lines indicating the dose profiles depicted in subfigures (b~d). A good match between oclMC results and gDPM results was also observed for this challenging case.

Quantitative comparison results for these photon cases were also presented in Table 1. The average relative dose difference within 10% isodose line was 0.15%, 0.16% and 0.17% for the water phantom, the slab phantom and the half slab phantom, respectively. More than 98% of the calculation points have a dose difference less than 0.53%, 0.67% and 0.73% for these three phantoms.

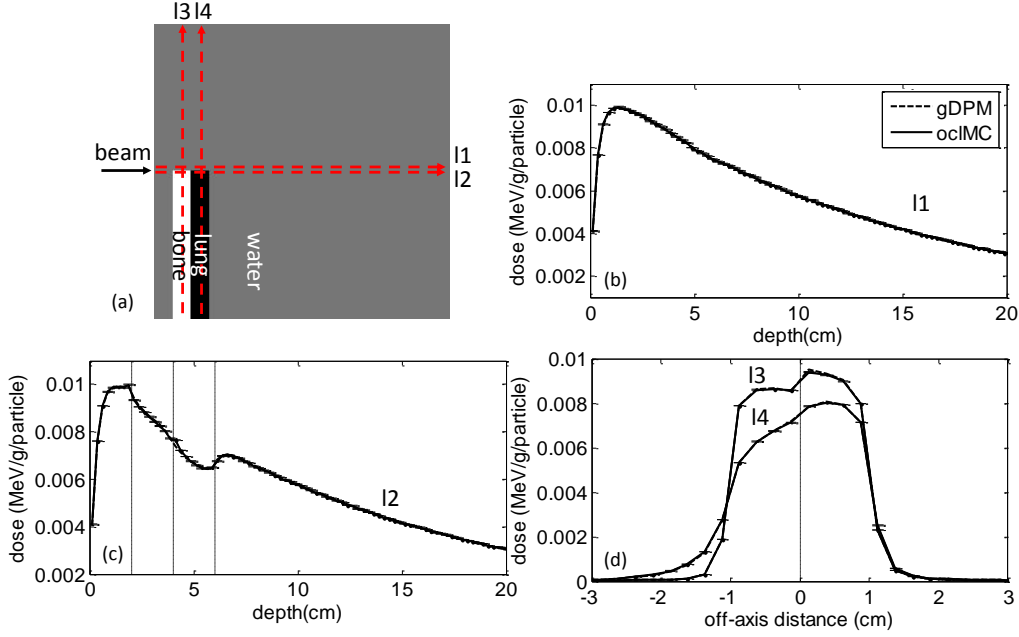

Figure 3. (a) Configuration of a half-slab phantom geometry. Dash lines indicate where dose profiles are plotted. (b)-(c) Depth dose along l1 and l2. (d) Lateral profiles along l3 and l4. Error bars show $2\sigma_D$ of dose calculated by our OpenCL-based dose engine.

## 3.2 Efficiency Evaluations

We first compared the computational efficiency of gDPM and oclMC running on the same Nvidia GTX TITAN GPU card. The results were summarized in Table 2. The computation time for oclMC was slightly longer than that for gDPM. This difference can be ascribed to both different physics transport models in these two packages and the different computational environment between CUDA and OpenCL. The time difference is about 14-17% for the electron cases and 6-17% for the photon cases. However, practically, both of the two dose packages have achieved very high computational efficiency and the difference between them is probably not significant for clinical applications.

**Table 2.** Computation time for different cases of gDPM and ?? on Nvidia GPU.

| Beam | No. of particles | Phantom | gDPM (s) Nvidia GeForce GTX TITAN | oclMC (s) Nvidia GeForce GTX TITAN | Ratio gclMC/gDPM |
|---|---|---|---|---|---|
| 15MeV electron | $5\times10^6$ | Water | 3.00 | 3.43 | 1.14 |
| | | Slab | 3.73 | 4.35 | 1.17 |
| 6MV photon | $5\times10^8$ | Water | 33.11 | 35.07 | 1.06 |
| | | Slab | 43.44 | 51.00 | 1.17 |
| | | Half Slab | 42.33 | 47.89 | 1.13 |

The gDPM package was developed on the Nvidia CUDA platform, which hence can only run on Nvidia's GPU. In contrast, our new dose package oclMC developed under OpenCL environment can run on a set of different platforms. To test its cross-platform portability, we ran oclMC on an Nvidia GTX TITAN GPU card, an AMD

Radeon R9 290x GPU card, an AMD HD 7500S GPU card, as well as an Intel i7-3770 CPU card using only one thread and using eight threads with four cores. We recorded the simulation time on these platforms, which were listed in Table 3. Comparing the computation time for the same code oclMC running on the Nvidia TITAN GPU and AMD Radeon 290x GPU, it was found that the AMD card was ~1.34 faster for the photon cases. The time difference for the electron cases was negligible. Comparing the hardware specifications of these two GPU cards, we found very similar numbers, namely 2688 vs 2816 stream processing units, 876 vs 980 MHz GPU clock speed, and 288.4 vs 352 GB/s memory bandwidth. However, the AMD Radeon 290x card has much more multiprocessors (MP) with fewer stream processing units per (i.e., 44 MP with 64 processing units on each MP) than Nvidia TITAN card (14 MP with 192 units per MP). This AMD configuration should help to alleviate the GPU thread divergence issue, a major issue for GPU-based MC simulation. Specifically, although NVidia card groups more stream processing units to a MP, the thread divergence issue hinders parallel processing capability among these units, reducing the effective parallel processing power. Hence, our oclMC code experienced efficiency gain when moving from the Nvidia TITAN card to the AMD 290x card.

The computation time of our oclMC code on the AMD 7500S GPU was much longer than those of the other two GPU cards mentioned above. This is due to its much lower hardware processing power, e.g. 6 MP with a total number of 384 stream processing units, 650 MHz GPU clock speed, 72 GB/s memory bandwidth. In addition, in our test run this AMD 7500S GPU card was also responsible for display of a desktop, which further reduces the computational efficiency.

**Table 3.** Computation time for different cases of oclMC on different platforms.

| Beam | No. of particles | Phantom | oclMC (s) | | | | |
|------|------------------|---------|-----------|---|---|---|---|
| | | | Nvidia GeForce GTX TITAN | AMD Radeon R9 290x | AMD Radeon HD 7500S | Intel i7-3770 CPU (4 cores, 8 threads) | Intel i7-3770 CPU (single thread) |
| 15MeV electron | $5\times10^6$ | Water | 3.43 | 3.70 | 175.30 | 50.52 | 260.01 |
| | | Slab | 4.35 | 4.75 | 182.70 | 54.01 | 296.57 |
| 6MV photon | $5\times10^8$ | Water | 35.07 | 28.47 | 1461.18 | 473.87 | 2672.85 |
| | | Slab | 51.00 | 35.57 | 1840.16 | 529.62 | 2955.84 |
| | | Half Slab | 47.89 | 35.54 | 1783.42 | 525.48 | 2930.57 |

Comparing the computation time running on the Nvidia GPU and the CPU (Intel i7-3770@3.4GHZ), it was found that the speedup factor of the Nvidia GPU is about 12.42-14.73 times for electron cases and 10.38-13.51 times for photon cases. These numbers considered the parallel processing on this i7 processor using all the 4 cores and 8 threads. We also ran our code on the same i7 processor using only 1 thread, which slowed down the performance on CPU side by about 5.15-5.5 times and 5.58-5.64 times for the electron cases and the photon cases, respectively.

In addition, it was also noted that in both the electron and the photon cases, the computation times for the heterogeneous geometries were longer than those for the homogeneous phantom. For the electron cases, this was due to the fact that the mean free path of electrons in bone was much shorter than that in water and thus shortened sampled steps for electron transport in bone. Although the mean free path of electrons in lung was much longer than that in water, the length of the sampled step in lung was mainly limited by the user-specified maximal step size and voxel size. As for the photon case, the time difference was mainly caused by the Woodcock method. In a heterogeneous phantom, this Woodcock method transported photons as if in a homogeneous phantom that had the highest attenuation coefficient found in the heterogeneous phantom. It then sampled a lot of fictitious interactions to account for the difference between this highly attenuating homogeneous phantom and the actual heterogeneous one. In the cases studied here, the algorithm transported photons in the slab and half slab geometries as if in a homogeneous bone phantom. This yielded a lot more interactions sampled than in the water case, most of which were fictitious interactions, which slowed down the computations.

## 4. Discussions and Conclusion

In this paper, we have presented our new fast Monte Carlo dose calculation package oclMC for radiotherapy, developed under OpenCL cross-platform environment. The accuracy of oclMC was benchmarked against our previously developed CUDA-based package gDPM. Satisfactory agreements have been observed for both electron and photon beams on homogenous water phantoms and heterogeneous slab and half-slab phantoms, indicating the accuracy of our package. The simulation times of our new dose package for different cases were comparable to those of gDPM on an Nvidia TITAN GPU card. The cross-platform portability of oclMC was demonstrated by running it on different platforms, including a Nvidia GeForce GTX TITAN GPU card, an AMD Radeon R9 290x GPU card, an AMD Radeon HD 7500s GPU card, as well as an Intel i7-3770 CPU card. To our knowledge, this is the first MC package that was developed under the OpenCL environment with cross-platform capability.

It was noticed that there were small dose discrepancies between oclMC and gDPM calculated dose distributions around the interface of different materials for the slab phantom cases. This was due to different schemes to handle the electron boundary crossing issues in these two MC dose engines. Specifically, gDPM employed a fixed step algorithm, where an electron can cross multiple voxel boundaries before a hard event occurs. This was made possible by a scaling function of cross section that approximately holds in the radiotherapy energy regime (Sempau *et al.*, 2000). In contrast, oclMC limited an electron transport step to be within a voxel. However, these small dose discrepancies were probably acceptable for clinical applications.

The computational speed of this new code oclMC was found to be slightly slower than that of gDPM. This is the consequence of a number of different factors. The fixed step algorithm adopted in gDPM allows an electron to cross multiple voxel boundaries before a hard event occurs, which makes gDPM more computationally efficient. This was actually the case for the CPU platform. However, when it comes to the GPU platform, the frequent voxel boundary crossing in one electron step increases the possibility of GPU thread divergence. Hence, the advantages of gDPM on the algorithm side might be counteracted by the GPU platform to some extent. The electron stepping algorithm used in oclMC was also found to have a significant impact on the computation efficiency. On one hand, since an electron step is always limited to be within a voxel, the computation speed becomes heavily dependent on the voxel size. In our experiments, we found that ~2.5 mm voxel size can provide both clinical acceptable resolution and comparable simulation efficiency to gDPM. On the other hand, the stepping algorithm makes different GPU threads better synchronized with each other and is thus GPU-friendly in principle.

One definite advantage of our OpenCL-based MC dose engine oclMC is its cross-platform portability, as demonstrated by the feasibility of running it on CPU and GPUs from Nvidia and AMD. In the present form of oclMC, the kernel codes do not need to be modified to run on different platforms, although the peripheral part is platform specific to configure the execution environment. Keeping this portability prevents tailoring our dose engine down to the very specific platform, which is hence not advantageous for speed consideration. However, the overall performance is only ~15% lower than gDPM in all the cases tested. This is probably acceptable for practical applications, especially considering the achieved dose calculation time on one NVidia GPU card is only seconds to tens of seconds. Moreover, running our new code on the AMD 290x card yields much shorter absolute processing time than running gDPM on an NVidia Titan card because of the stream processing unit configuration of the AMD card that is favored by MC simulations. This high efficiency is an advantage of our new code.

**Acknowledgements**

This work was supported by ??